\documentclass[twocolumn,showpacs]{revtex4}
\usepackage[xdvi]{graphicx}%
\usepackage{dcolumn}
\usepackage{amsmath}
\makeatletter

\def\btt#1{\texttt{\@backslashchar#1}}%
\DeclareRobustCommand\bblash{\btt{\@backslashchar}}%
\makeatother

\newcommand{\bra}{\left\langle}
\newcommand{\ket}{\right\rangle}
\newcommand{\kets}[1]{\right\rangle^{{\rm s}}_{#1}}

\newcommand{\keth}[1]{\right\rangle_{#1}}

\newcommand{\pder}[2]{\frac{\partial #1}{\partial  #2}}

\newcommand{\der}[2]{\frac{d #1}{d  #2}}

\newcommand{\ps}[1]{p^{{\rm s}}_{#1}}

\newcommand{\e}{{\rm e}}

\begin{document}

\title{Extended Einstein relations with a complex effective 
temperature in a one dimensional driven lattice gas}
\author{Kumiko Hayashi and Shin-ichi Sasa}
\affiliation
%\address
{Department of Pure and Applied Sciences,  
University of Tokyo, Komaba, Tokyo 153-8902, Japan}

\date{\today}

\begin{abstract} 
We carried out numerical experiments on a one-dimensional driven
lattice gas to elucidate the statistical   properties of steady 
states far from equilibrium.  By   measuring the  bulk density 
diffusion constant $D$, the conductivity  $\sigma$, the intensity 
of density fluctuations $\chi$, we confirm that the Einstein 
relation $D\chi=\sigma T$, which is valid in the linear response 
regime about equilibrium, does not hold in such steady states.  
Here, $T$ is the environment  temperature,   and the Boltzmann 
constant is set to unity. Recalling that the Einstein  relation 
provided the first step in the construction of linear  response 
theory, we attempt to extend it to a generalized form valid in 
steady states far from equilibrium. In order to obtain new relations 
among measurable quantities,    we  define a  complex  effective 
temperature $\Theta-i\Phi$ from studying the static response of the 
system to a slowly varying potential in space. Replacing $T$ in the 
Einstein relation by the real part of the effective temperature 
$\Theta$, we     numerically    confirm   that   the   relation 
$D\chi=\sigma\Theta$ holds in the nonequilibrium steady states far 
from equilibrium that we study.  In addition to this extended form, 
we find the new relation $(L/2\pi)c\chi=\sigma\Phi$, where $c$ 
represents the propagation velocity of density fluctuations. 
\end{abstract}

\maketitle

%%%
%%%
\section{introduction}
%%%
%%%

%%% Einstein relation %%%

In 1905, Einstein derived the relation $D=\mu T$ between the 
diffusion constant $D$ for non-interacting Brownian particles 
suspended in an equilibrium fluid of temperature $T$ and mobility 
$\mu$ with respect to an  external   force \cite{ein}. (Here, the 
Boltzmann constant is set to unity.)  This relation is now called 
the 'Einstein relation'. Through the development of linear response 
theory, the original Einstein relation was extended to the form
\begin{equation}
D \chi=\sigma T, 
\label{eqein}
\end{equation}
which is valid  for interacting many-body systems \cite{ein2}. 
Here, $D$ should be considered as  the bulk density diffusion 
constant, $\chi$ is the intensity of density fluctuations, and 
$\sigma$ is the conductivity.  

%%% motivation %%%

The Einstein relation is  a universal relation in the linear 
response regime about equilibrium. It represents the first step 
in the construction of nonequilibrium statistical mechanics in 
the linear response regime.  With this in mind, if the  Einstein 
relation can be  extended to some form valid for nonequilibrium 
steady   states (NESS)  far from   equilibrium,   realizing an 
understanding of this form will   be an  important step in the 
construction of   statistical   mechanical   theory of   NESS.

%%% specific question %%%

With this motivation, because  it is simpler to study one-particle 
NESS systems than many-body systems, we previously analyzed a 
one-particle Langevin equation with a tilted periodic potential 
\cite{HSIIIa}. In that study, it was found that there exists an 
extended Einstein relation  with an effective temperature that 
can be measured independently of the diffusion constant and the 
differential mobility. Then, it is natural to ask whether that 
result for a   Brownian   particle   in NESS can be extended to 
many-body interacting systems.  In the present paper, we address 
this question by studying a one-dimensional driven lattice gas (DLG) 
\cite{KLS, BKS, SP} as an example of a many-body interacting system.

%%% 
%%%
\section{Model}
%%%
%%%

Let $\eta_x$ be an occupation variable defined on each site $x$ 
of a  one-dimensional  periodic  lattice $x=1,2, \cdots, L$. 
The variable $\eta_x$ takes the value $1$ if the $x$-th site is 
occupied by a particle  and $0$ if it is empty. In   order   to 
describe the model simply, we introduce the variable $\eta_{L+1}$, 
which is assumed to be equal to $\eta_1$. The array of occupation 
variables $(\eta_1, \cdots, \eta_L)$ is  denoted by  $\eta$ and  
called the ``configuration''.   

The time evolution of $\eta$ is described by the following rule: 
At each time step, randomly choose a nearest neighbor pair $x$ 
and $x+1$, where $1 \le x \le L$,  and  exchange  the  values  of 
$\eta_x$ and $\eta_{x+1}$ with   probability $c(x;\eta)$  given by  
$c(x;\eta)=\alpha \e^{-\frac{1}{2}\beta Q(\eta\to\eta^x)}$, where 
$\eta^{x}$ is the configuration obtained from $\eta$ through this 
exchange, $\beta=1/T$ is the inverse temperature of the environment, 
and $Q(\eta \to \eta^{x})$ represents the heat absorbed from the 
heat bath as a result of change of configuration $\eta\to\eta^{x}$ 
in one Monte Carlo step. The constant $\alpha$ is determined so 
that $c(x;\eta) \le 1$. The particle number $N=\sum_{x=1}^L\eta_x$ 
is  conserved throughout the  time  evolution,  and  the  averaged 
density $\bar \rho=N/L$ is a parameter of the model. 

In this paper, we study the DLG for which  
\begin{equation}
Q( \eta \to \eta')
= H_0( \eta' )-H_0(\eta) - E \sum_{x=1}^L j_x(\eta \to \eta') ,
\label{derh}
\end{equation}
where $E$ is a constant driving   force,   $H_0(\eta)$  is  the 
equilibrium Hamiltonian $H_0(\eta)= -\sum_{x=1}^L\eta_x\eta_{x+1}$,  
and $j_x(\eta \to \eta')$ is the net number of particles hopping  
from $x$ to $x+1$  for  the  configuration  change  $\eta\to\eta'$ 
in one Monte Carlo step. In the argument below,  $\bra \  \kets{E}$ 
represents the statistical average in a NESS with driving  force $E$. 

Throughout this paper, we fix $\beta=0.5$,  and the numerical 
parameter $\alpha$  introduced  in  the  exchange  probability 
$c(x;\eta)$  is set  to $\exp(-3\beta)$. This value is chosen  so 
that $c(x,\eta)\le 1$ in the case $E \le 5$. As a typical example 
of a NESS far from equilibrium, we study this model with  $E=4$, 
and we consider the cases  $\bar\rho=19/64$ and $0.5$, with $L=64$, 
$128$ and $256$.  We remark that because we attempted to evaluate 
precisely  measurable quantities introduced in the next section, 
it was hard to obtain data in the case $L > 256$. 

\begin{figure}
\begin{center}
\includegraphics[width=8cm]{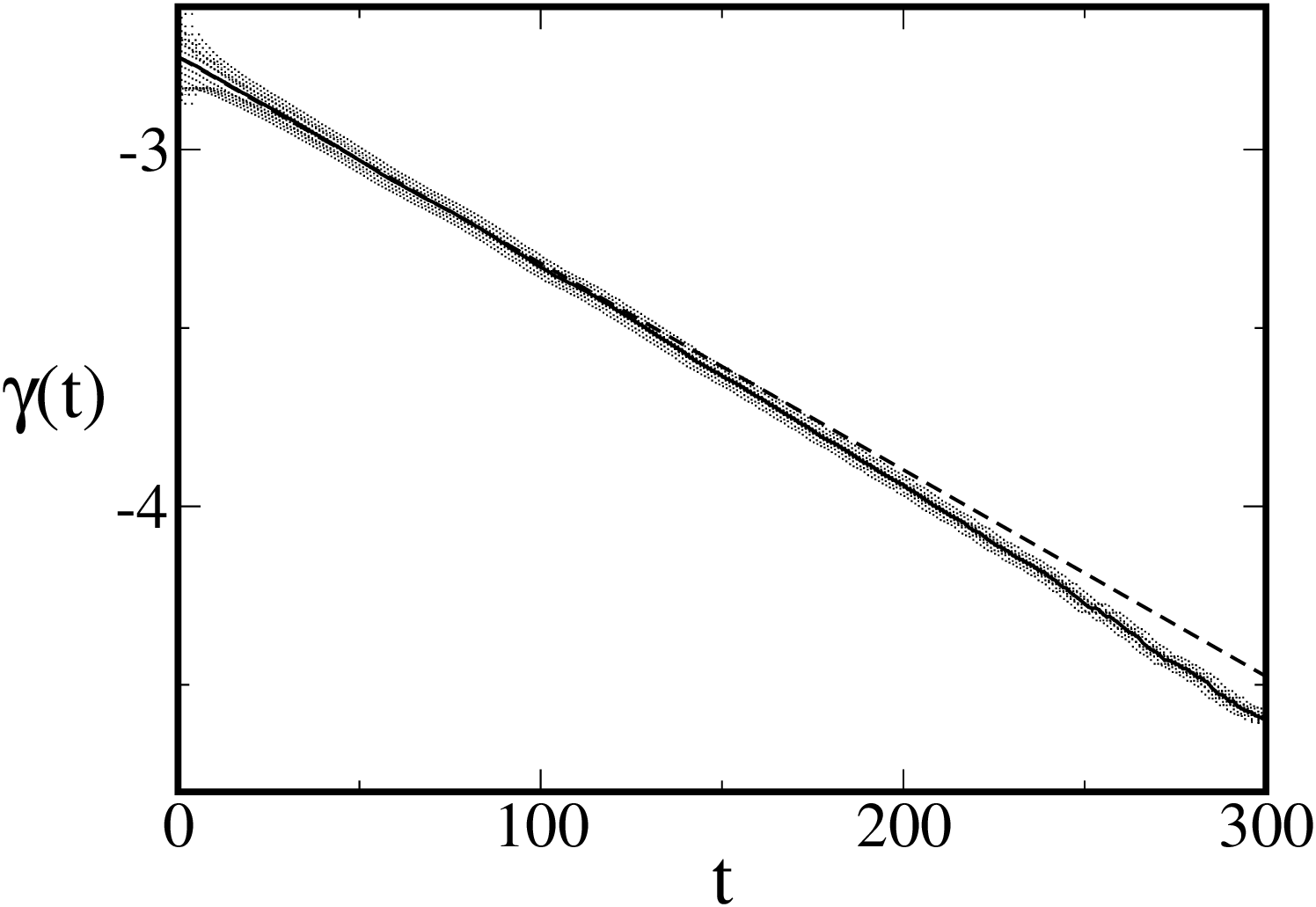}
\end{center}
\caption{$\gamma(t)$  is plotted  with error bars  in the case   
$\bar\rho=0.5$, $L=64$ and $|\hat V_1|/\sqrt{L}=0.1$ (See 
(\ref{d:diff})). The slope of the fitted line (the dotted line) 
in the early stage ($t <  150$ mcs)  corresponds to $D(2\pi/L)^2$.  
The slope of  $\gamma(t)$  in  the late stage ($t > 150$ mcs) 
differs from  $D(2\pi/L)^2$. 
}  
\label{fig:gamma}
\end{figure}
%
%

%%%
%%%
\section{Einstein relation} 
%%%
%%%

Because we have little knowledge regarding relations characterizing  
NESS of DLG, let us start by investigating the  validity of the 
Einstein relation (\ref{eqein}) by defining the quantities $\sigma$, 
$\chi$ and $D$ for this model.  First,  using the averaged current 
$\bar J_E$, which  is defined by  
\begin{equation}
\bar J_E= \frac{1}{L}\bra  \sum_{x=1}^L j_x(\eta \to \eta') \kets{E},
\end{equation}   
the conductivity $\sigma$ is expressed as  
\begin{equation}
\sigma=  \frac{d \bar J_E}{dE}. 
\label{sigdef}
\end{equation}
Second, the intensity of large scale density fluctuations is defined 
by 
\begin{equation}
\chi (L)= \bra  |\hat \eta_1|^2 \kets{E},
\label{chiL}
\end{equation}
where 
\begin{equation}
\hat \eta_n(t) =  \frac{1}{\sqrt{L}}
\sum_{x=1}^{L} \eta_x(Lt) \e^{ i\frac{  2\pi n }{L}x }.
\label{hedef} 
\end{equation}  
Third, the density  diffusion  constant  $D$  is  defined  as  the 
coefficient of  the  diffusion  term  of  the  evolution  equation 
describing the  averaged  behaviors  of  the density field.  As an 
example of such  behaviors, we consider the time  evolution  of 
the density field after a slowly varying weak potential $V_x$ is 
turned on at $t=0$ in the case that the  system  is  in  a  steady 
state. Here $V_x$ is given by
\begin{equation}
V_x=\frac{1}{\sqrt{L}} \hat V_1\e^{-i\frac{2\pi}{L} x}+\rm c.c, 
\label{pote}
\end{equation}
where c.c denotes the complex conjugate, and $\hat V_1$ is a 
complex  number  satisfying  $|\hat V_1| /\sqrt{L}\ll 1$. 

\begin{figure}
\begin{center}
\includegraphics[width=8cm]{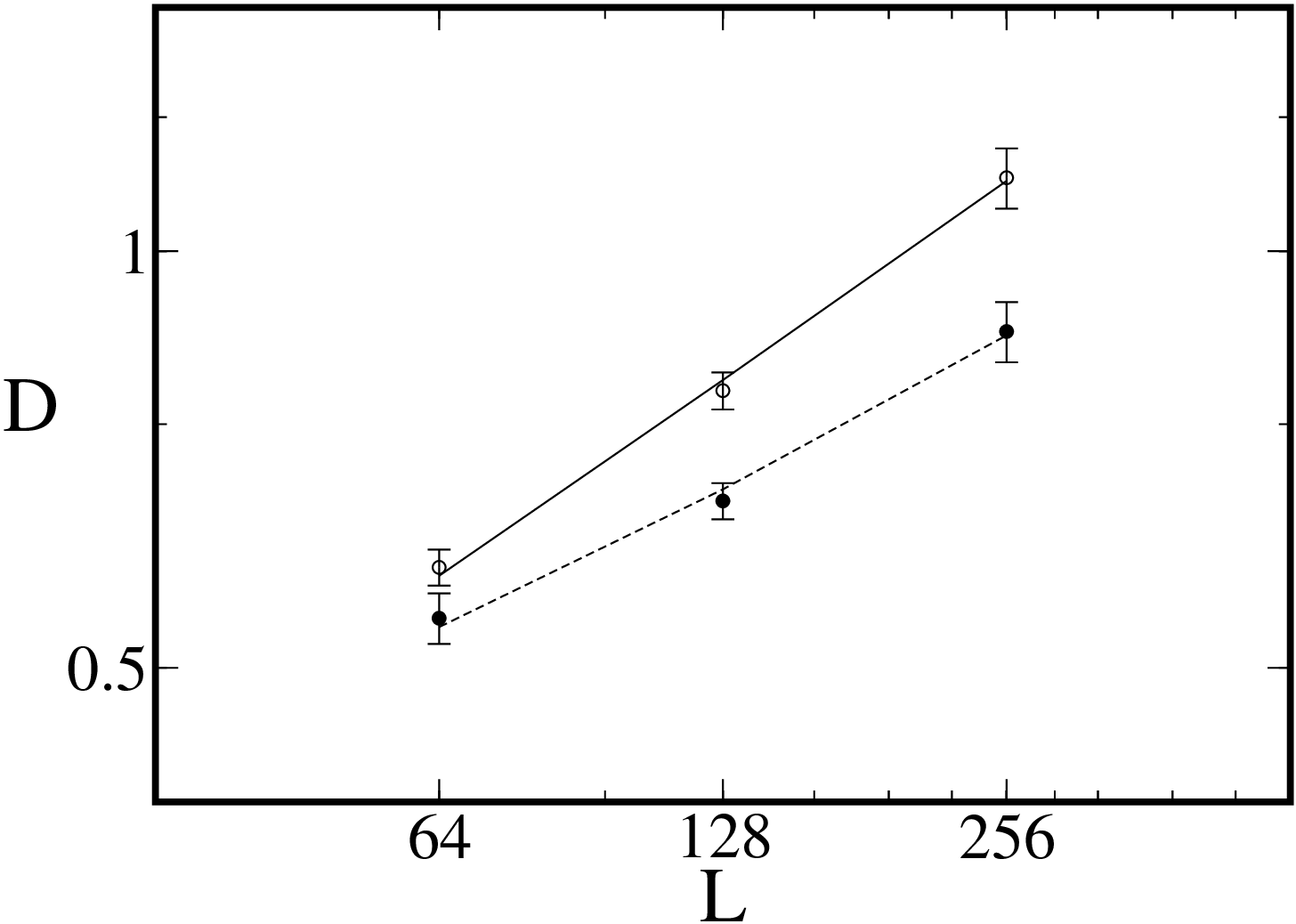}
\end{center}
\caption{$D$  as a function of $L$ for the case $\bar\rho=19/64$ 
(solid circles) and $\bar\rho=0.5$ (open circles).  The dotted 
line  and the solid  line  represent  $0.042\sqrt{L}+0.20$  and 
$0.068\sqrt{L}+0.0413$, respectively.    
}
\label{fig:dif}
\end{figure}  

The averaged behavior  in this case is represented approximately  
by the time evolution of a quantity $\bra \hat\eta_1(t)\keth{E}^
{V}$,  where $\bra \ \keth{E}^{V}$ denotes the statistical average 
under the condition that the equilibrium Hamiltonian becomes 
$H_{V}(\eta)=H_0(\eta)+ \sum_{x =1}^{L}\eta_x V_x $ when $t >0$. 
We define the response function against $V_x$ as 
\begin{equation}
R(t)=- \lim_{|\hat V_1| \to 0}
\frac{\bra\hat \eta_1(t)\ket_E^V}{\hat V_1}. 
\label{res}
\end{equation}  
In order to see the behavior of $R(t)$, in Fig. \ref{fig:gamma}, 
we plot 
\begin{equation}
\gamma(t)= \log \left|-\frac{\bra\hat \eta_1(\infty)\ket_E^V}
{\hat V_1} +\frac{\bra\hat \eta_1(t)\ket_E^V}{\hat V_1} \right|. 
\label{gamma} 
\end{equation}
Note that we have  checked that $\gamma(t)$ takes the same value 
for the cases $|\hat V_1|/\sqrt{L}=0.05$, $0.1$, $0.15$ and $0.2$. 
It implies that we can estimate the value of the right hand  of  
(\ref{res}) within a numerical accuracy. It is seen in Fig. 
\ref{fig:gamma} that $\gamma(t)$ has the linear slope in the 
early time regime. Therefore, using this linear slope, $D$ is 
estimated from the form 
\begin{equation} 
|R(\infty)-R(t)| 
= {\rm const.}\e^{- D \left(\frac{2\pi}{L}\right)^2 t},
\label{d:diff}
\end{equation}
in the early time regime. 
(We remark that we cannot deny a possibility that the linear 
slope shown in the early time regime might be caused by finite size 
effects.  This is an unresolved problem.)

\begin{figure}
\begin{center}
\includegraphics[width=8cm]{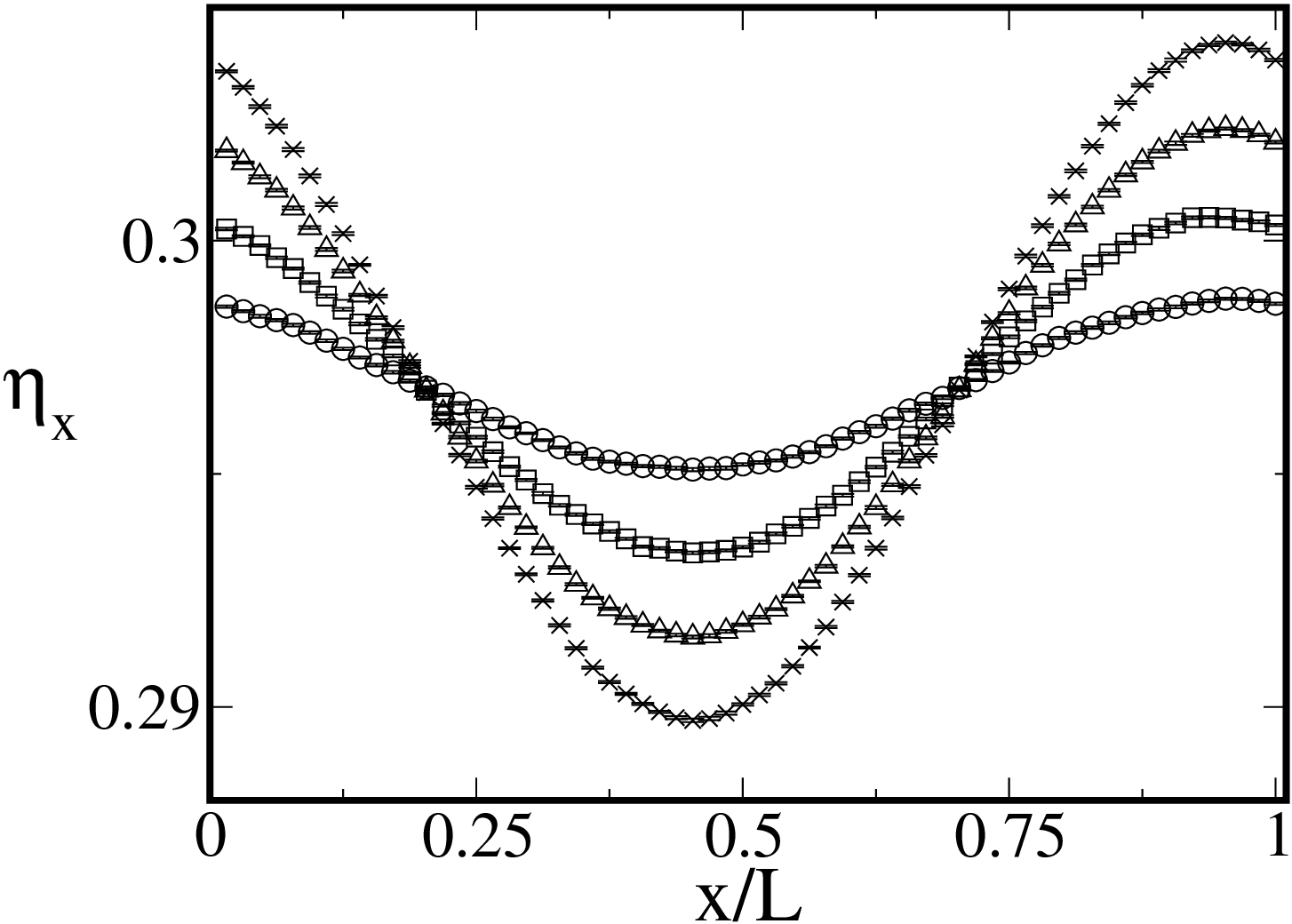}
\end{center}
\caption{The profiles of $\bra\eta_x\kets{E,V}$ under the influence 
of $V_x=\Delta\sin(2\pi x/L)$  in  the  case  $\bar\rho=19/64$ and 
$L=64$.  $\Delta=0.1$, $0.2$, $0.3$ and $0.4$ correspond to the 
circle, square,  triangle and plus, respectively.     
}
\label{fig:conf}
\end{figure}

Using the above definitions of $\sigma$, $\chi$ and  $D$, we 
confirm the validity  of  (\ref{eqein})  numerically  for  the 
equilibrium case, $E=0$. With regard to other  relations in the 
linear  response  theory,  we can derive (\ref{eqein})  from the 
Green-Kubo relation and the fluctuation response relation, which  
have  been  proved  for  DLG  models \cite{HSIVb}. Contrastingly, 
in NESS for our DLG, we find that the two sides of (\ref{eqein}) 
differs greatly.  Specifically, in the case $\bar\rho=19/64$, we  
have  confirmed  that  $\sigma(L)$  and  $\chi(L)$  converge  to 
$\sigma_*=0.032$  and  $\chi_*=0.26$ in the limit $L \to \infty$, 
while  $D$  has the anomalous $L$ dependence $D(L)\propto\sqrt{L}$,   
as  shown  in  Fig. \ref{fig:dif}.  This anomalous behavior  is  
essentially  the same as that reported in  Ref. \cite{BKS}. Also, 
in  the  case  $\bar\rho=0.5$, (\ref{eqein}) is not valid because 
we have confirmed that $\sigma_*=0.037$  and  $\chi_*=0.32$  in   
the  limit     $L \to \infty$,   while  $D(L)\propto\sqrt{L}$   
(See Fig. \ref{fig:dif}).  

Now we ask whether it is possible to obtain an extension of 
the Einstein relation valid for NESS  of interacting many-body 
systems by defining an appropriate measurable effective temperature. 
This question arises naturally, because we have found that such an 
extended  Einstein  relation  does  hold  for  NESS of one-particle 
systems. In this relation, the temperature of the environment, $T$, 
is replaced by  an  effective  temperature, which can be measured 
independently of the other  quantities  appearing  in  the  Einstein 
relation \cite{HSIIIa}.  By analogy to that case, we seek to define 
an effective  temperature  which  when  used  in  place  of  $T$  in 
(\ref{eqein}) provides an extended Einstein relation that is valid 
for NESS in interacting many-body systems.

%%%
%%%
\section{Extended Einstein relations}\label{s:ein}
%%%
%%%

%
\begin{figure}
\begin{center}
\includegraphics[width=8cm]{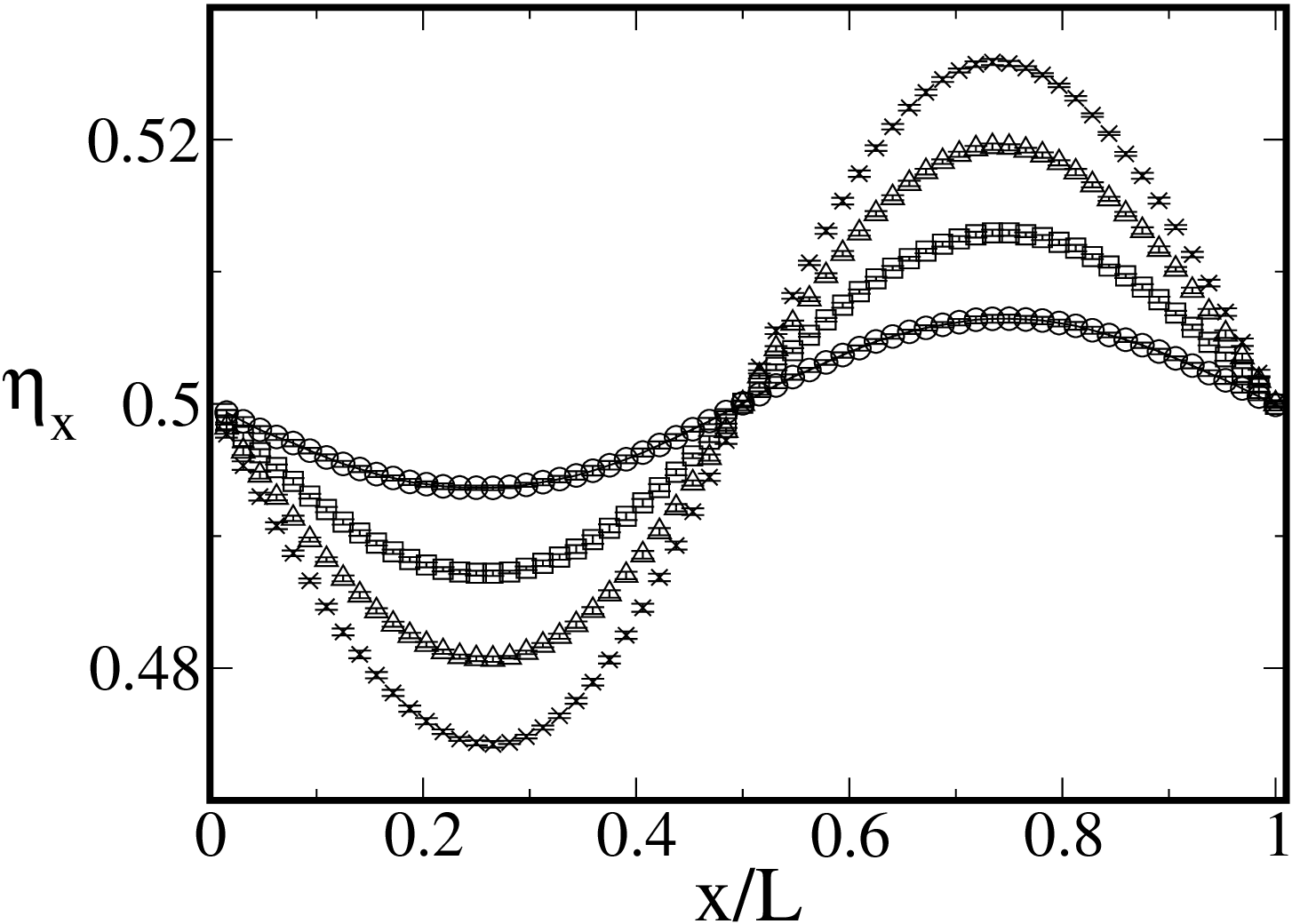}
\end{center}
\caption{The profiles of $\bra\eta_x\kets{E,V}$ under the influence  
of  $V_x=\Delta\sin(2\pi x/L)$  in  the  case $\bar\rho=0.5$ and 
$L=64$. $\Delta=0.1$, $0.2$, $0.3$ and $0.4$ correspond to the circle, 
square,  triangle and  plus, respectively.
}
\label{fig:conf2}
\end{figure}

\begin{figure}
\begin{center}
\includegraphics[width=8cm]{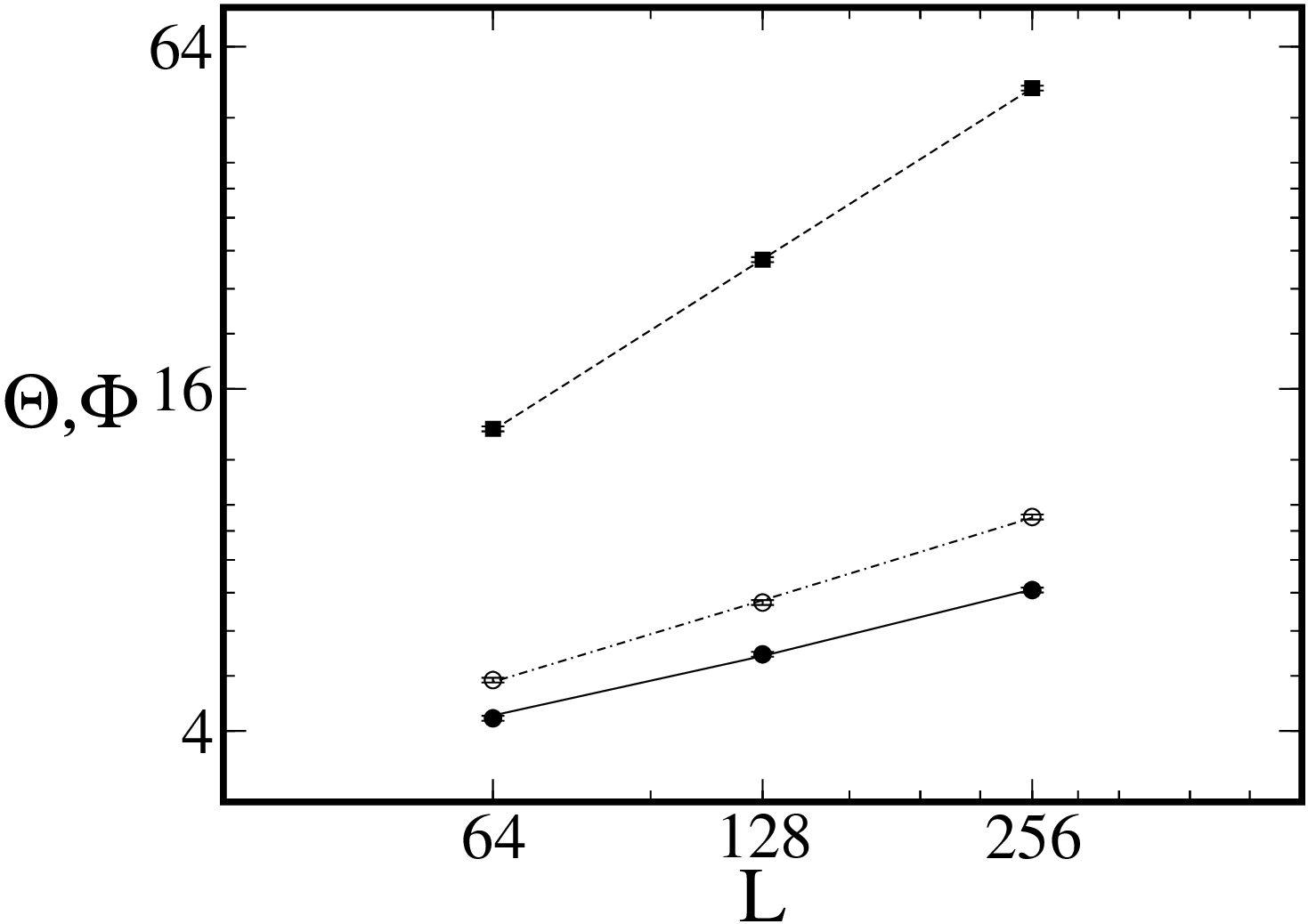}
\end{center}
\caption{$\Theta$ and $\Phi$ as  functions of $L$ in the cases  
$\bar\rho=19/64$ (solid symbols) and $\bar\rho=0.5$ (open symbols).  
$\Theta(L)$ and $\Phi(L)$ correspond to the circle and square, 
respectively.  The solid  line, the dotted line and the dash-dotted  
line represents  $0.36\sqrt{L}+1.4$, $0.21L$ and $0.58\sqrt{L}+0.26$. 
Note that $T=2$. 
}
\label{fig:the}
\end{figure}

In a previous study, we showed that the  effective temperature 
appearing in  the Einstein relation valid for NESS of a Brownian 
particle can be measured by adding a slowly varying potential to 
the system \cite{HSIIIa}. In a similar way, we wish to determine 
an effective temperature by adding to the system a slowly varying 
potential given by (\ref{pote}).  Due to  the influence of this 
potential, we have $\bra \hat \eta_1 \kets{E,V} \not = 0$, where  
$\bra \ \kets{E,V}$ denotes  the  steady  state  average  in  the 
system with driving force $E$ and potential $V_x$. Recalling that 
the relation  
\begin{equation}
\bra \hat \eta_1 \kets{E=0,V} = -\hat V_1 \frac{\chi}{T} +o(|\hat V_1|), 
\label{eqeq}
\end{equation}
holds  in   the  equilibrium case, we  define the 
effective temperature $\hat\theta_1$  in NESS as 
\begin{equation}
-\lim_{|\hat V_1| \to 0} \frac{\bra \hat \eta_1 \kets{E,V}}{\hat V_1}    
=\frac{\chi(L)}{\hat \theta_1}. 
\label{ctheta}
\end{equation}  
Note that in NESS, the left-hand side of (\ref{ctheta}) is a 
complex number in general. To see this concretely,  we show that 
$\bra\eta_x\kets{E,V}$ under the influence of the potential 
$V_x=\Delta\sin(2\pi x/L)$ has a component of $\cos(2\pi x/L)$  
(see Fig. \ref{fig:conf}). Expressing  explicitly  that 
$\hat\theta_1$ is a complex number, we write 
\begin{equation}
\hat \theta_1=\Theta - i \Phi,
\label{aa}
\end{equation}
where $\Theta$ and $\Phi$ are real numbers.  In Fig. \ref{fig:the},  
$\Theta$ and $\Phi$ are displayed as functions of $L$ in the case 
$\bar\rho=19/64$ and $\bar\rho=0.5$.  It is  seen  that  we  have  
the   anomalous  $L$  dependences  $\Theta(L)\propto\sqrt{L}$  and 
$\Phi(L)\propto L$. We remark that $\Phi=0$ in the case $\bar\rho=0.5$ 
(See Fig. \ref{fig:conf2}). 

%
%%
%%\section{Extended Einstein relations:}  
%%

%

\begin{figure}
\begin{center}
\includegraphics[width=8cm]{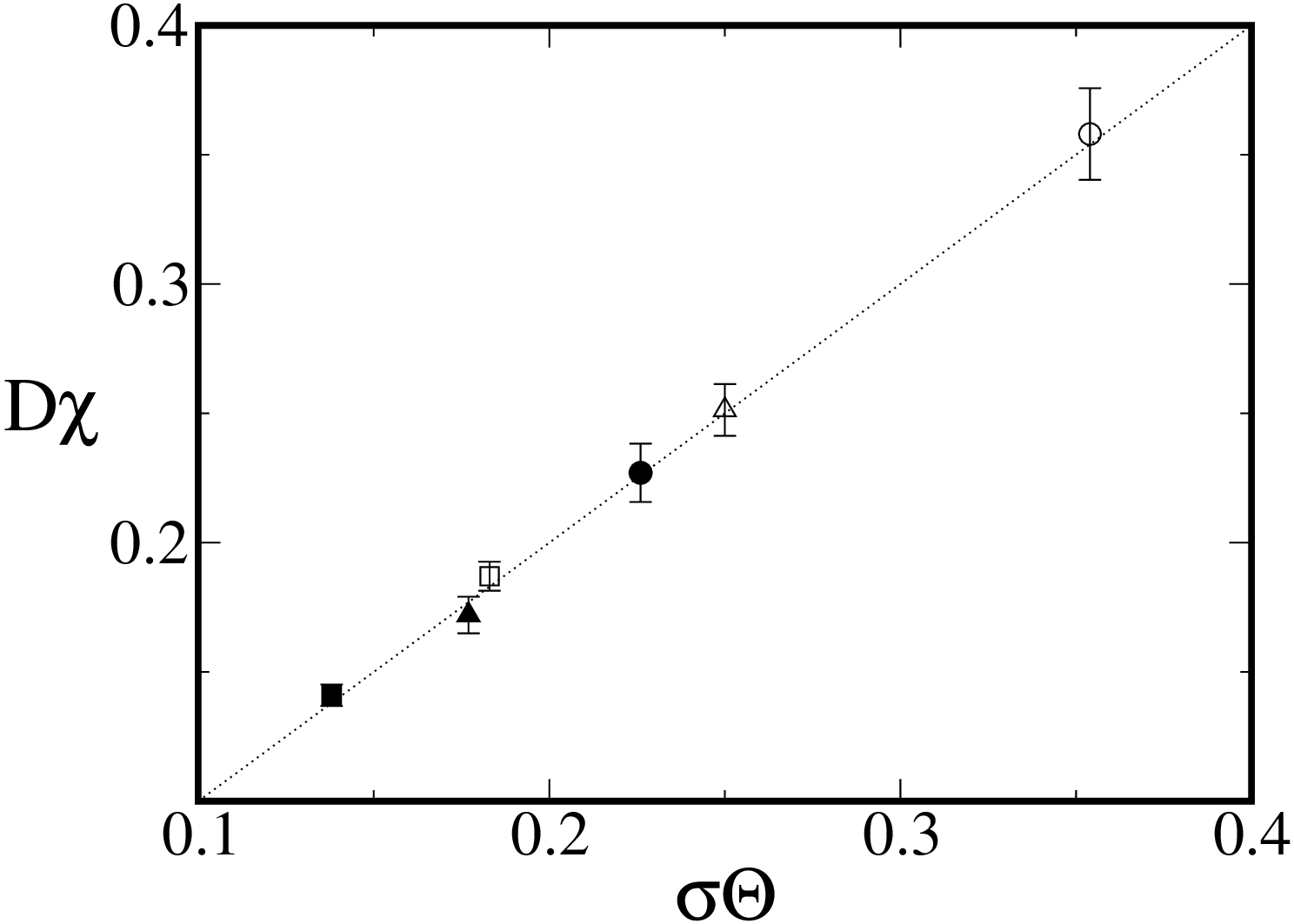}
\end{center}
\caption{$D\chi$  as a function of $\sigma\Theta$. The cases 
$\bar \rho=19/64$ and $\bar \rho=0.5$ correspond to the solid 
symbols and the open symbols, respectively. The cases $L=64,128$ 
and  $256$  correspond to the  square,  triangle, and  circle, 
respectively.  The straight line represents  $D\chi=\sigma\Theta$.   
}
\label{fig:ein1}
\end{figure}

Noting  that from the graphs in Fig. \ref{fig:dif} and Fig. 
\ref{fig:the},  $D(L)$  and  $\Theta(L)$ are both approximately 
proportional to   $\sqrt{L}$,   we investigate whether the 
Einstein  relation  can  be  extended  to  the  NESS  by  simply 
replacing $T$ in (\ref{eqein}) by $\Theta(L)$. In Fig. 
\ref{fig:ein1}, we plot $D(L) \chi(L)$ as a function of  
$\sigma(L)\Theta(L)$ for the cases $\bar\rho=19/64$ and $0.5$, 
with  $L=64, 128$  and  $256$. The  graph  clearly  shows  that  
the extended Einstein relation  
\begin{equation}
D(L)\chi(L)=\sigma(L)\Theta(L)
\label{ein1}
\end{equation} 
holds within the precision of the numerical computations \cite{ER}. 

%%%
%%%
\section{failure of  extended fluctuation  response relation}\label{s:frr}
%%%
%%%

There are universal relations in the linear response regime about 
equilibrium, such as the Einstein relation, the fluctuation-response 
relation (FRR) and the Green-Kubo relation. One may expect that other 
relations can be extended in the similar way as did for the Einstein 
relation (\ref{ein1}) in NESS far from equilibrium. In this section, 
we  investigate an extension of the FRR. 

The FRR connects the time dependent response function and  the time 
correlation function. In order to define the response function,  we  
consider  the   situation  that  the perturbation (\ref{pote}) is 
applied to the system when $t=0$. The response function $R(t)$ in 
this situation is defined by (\ref{res}). In the equilibrium case, 
we can prove the FRR 
\begin{equation}
R(t)=\frac{1}{T}C(t),
\label{frr}
\end{equation}
with the time correlation function  
\begin{equation}
C(t)=\frac{1}{2}\bra |\hat \eta_1(t)- \hat \eta_1(0)|^2\kets{E}.
\label{corre}
\end{equation}  

\begin{figure}
\begin{center}
\includegraphics[width=8cm]{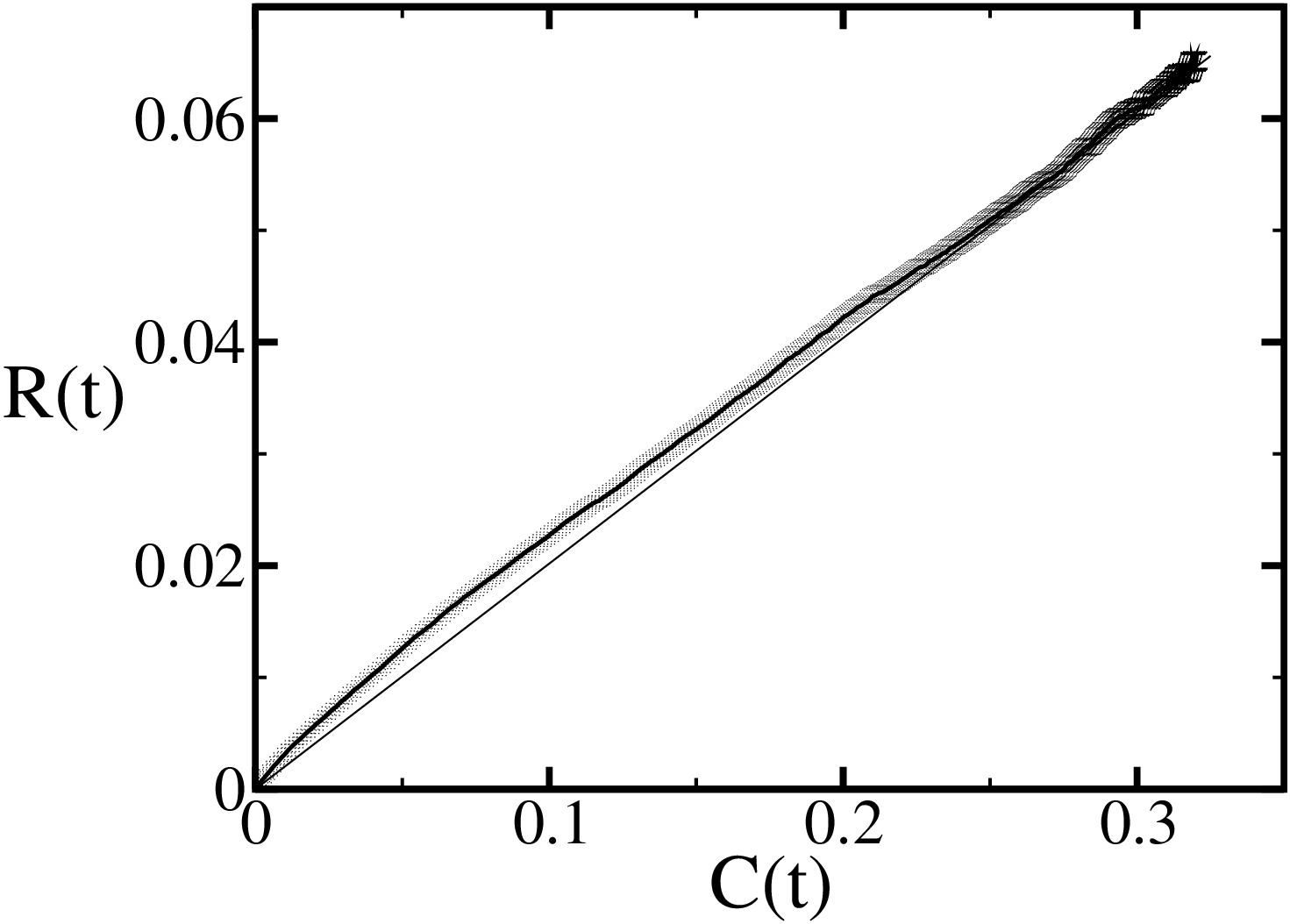}
\end{center}
\caption{$R(t)$ as a function of $C(t)$ in the case $\bar\rho=0.5$ 
and $L=64$ during an interval $0\le t\le 800$ mcs. The slope of the 
thin line represents $\Theta^{-1}$. 
}
\label{fig:frr}
\end{figure}

Because the Einstein relation was extended with the replacement of 
$T$ by $\hat\theta_1$, we expect that (\ref{frr}) can be extended 
in the form 
\begin{equation}
R(t)=\frac{1}{\hat\theta_1}C(t).
\label{frr2}
\end{equation}
Specifically,  we  numerically  investigate  the  validity  of  
(\ref{frr2})  in the case $\bar\rho=0.5$ and $L=64$. Because   
$\Phi=0$  in  the  case $\bar\rho=0.5$, (\ref{frr2}) is rewritten  
as  $R(t)=\Theta^{-1} C(t)$.   In  Fig. \ref{fig:frr}, we plot 
$R(t)$ as a function of $C(t)$ during an  interval $0\le t\le 800$ 
mcs.  It is seen that $R(t)$ deviates from $\Theta^{-1} C(t)$ 
substantially  in  an  early  time  regime ($0\le t\le 150$ mcs) 
where the diffusion constant $D$  was  defined (See (\ref{d:diff}) ). 
This implies that (\ref{frr2}) is not valid.  Note that $R(\infty)
=\hat\theta_1^{-1}C(\infty)$ holds by the very definition  of  
$\hat\theta_1$  given  by (\ref{ctheta}).

%%%  generalized version of FRR  %%%

We remark that a  generalized version of FRR of the nonlinear 
Langevin dynamics without the detailed balance condition was 
presented by several authors \cite{eyink}. However, in their 
formalism, a generalized force which defines a response function 
is chosen so that the simple form of  FRR holds, and it seems 
that their  response function cannot be  measured directly in 
experiments. In contrast to the previous studies, we consider a 
relation between $C(t)$ and $R(t)$ which are measured directly.  

%

%%%
%%%
\section{Interpretation}
%%%
%%%

%
%

%%% contents %%% 

In sections \ref{s:ein}  and  \ref{s:frr}, we find that the 
Einstein relation (\ref{ein1}) is extended with the diffusion 
constant $D$  which  is  measured  from $R(t)$ in the  early  
time  regime  and  the  effective  temperature $\Theta$, while 
the extended FRR  (\ref{frr2})  is  not  valid  unfortunately.  
In  paragraphs below,  we  shall provide  the phenomenological 
understanding of validity of (\ref{ein1}) and invalidity of 
(\ref{frr2}).  

%%% Einstein relation %%%

The extended Einstein relation  (\ref{ein1}) can be explained 
from the following  argument. Let $\bar J(\rho,E)$  be the 
averaged particle current in a system  with density $\rho$ and  
driving force $E$.    When the slowly varying potential (\ref{pote})  
is applied to this system, first   the   current   should   become  
$\bar J(\bra \eta_x \kets{E,V},E-dV_x/dx)$,  and  as  a  result  a 
diffusive current $ D \bra (\eta_{x+1}-\eta_{x}) \kets{E,V}$ should 
appear so as to make the  total current homogeneous in $x$; that is,  
it is reasonable to assume the quantity  
\begin{equation}
\bar J\left(\bra \eta_x \kets{E,V},E-\der{V_x}{x}\right)
-D \bra (\eta_{x+1}-\eta_x) \kets{E,V}
\label{balance}
\end{equation}
is independent of  $x$ in a  steady state. This can be rewritten as  
\begin{equation}
c \bra \hat \eta_1 \kets{E,V}
+i \sigma  \frac{2 \pi}{L} \hat V_1 
+i \frac{2 \pi}{L} D  \bra \hat \eta_1 \kets{E,V}=
o\left(\frac{|\hat V_1|}{L}, \frac{\bra \hat \eta_1 \kets{E,V}}{L}\right), 
\label{ein22}
\end{equation}
where $c=\partial\bar J(\bar\rho,E)/\partial\bar\rho$ represents the  
propagation velocity of   density fluctuations. {}From (\ref{ctheta}) 
and (\ref{ein22}), we obtain
\begin{equation}
\sigma \hat \theta_1=D \chi-i\frac{L}{2 \pi}\chi c.
\label{ein33}
\end{equation}
This leads to the extended Einstein relation (\ref{ein1}).

Here, we  note that a microscopic derivation 
of (\ref{balance}) is quite difficult. In particular, we have not 
been able to  show  that  $D$  defined  by  (\ref{d:diff})  and  
$D$  in (\ref{balance}) are identical. In relation to this point, we  
note that a different value of $D$ is obtained if  we consider the  
time regime $t \ge 150$ mcs in Fig. \ref{fig:gamma}. Using this 
value of  $D$,  the  relation  (\ref{ein1})  cannot  be  confirmed 
numerically. This result suggests that memory effects  in  a  single  
mode  description cannot be neglected in the late stage of  the 
relaxation process after the potential is turned on. Because the 
memory effects arise from a  nonlinear coupling between different 
modes, the simple assumption (\ref{balance}) in NESS is far from 
the trivial.  

%%% failure of FRR %%%

The non-trivial nature of the extended Einstein relation 
(\ref{ein1}) can be seen through the failure of the extended FRR 
(\ref{frr2}).  For example, because  we cannot observe the 
exponential decaying regime of $C(\infty)-C(t)$ even in the early 
time regime unlike the behavior of $R(\infty)-R(t)$ (See 
Fig. \ref{fig:gamma}), we cannot define the diffusion constant $D$ 
properly from the assumption that $\log|C(\infty)-C(t)|=- D 
\left(\frac{2\pi}{L}\right)^2 t+{\rm const.}$,  though in the 
equilibrium case, the same  value  of  $D$  is   obtained  whether  
it  is  defined  from  $\log|C(\infty)-C(t)|$ or 
$\log|R(\infty)-R(t)|$  because of the FRR (\ref{frr}). 

\begin{figure}
\begin{center}
\includegraphics[width=8cm]{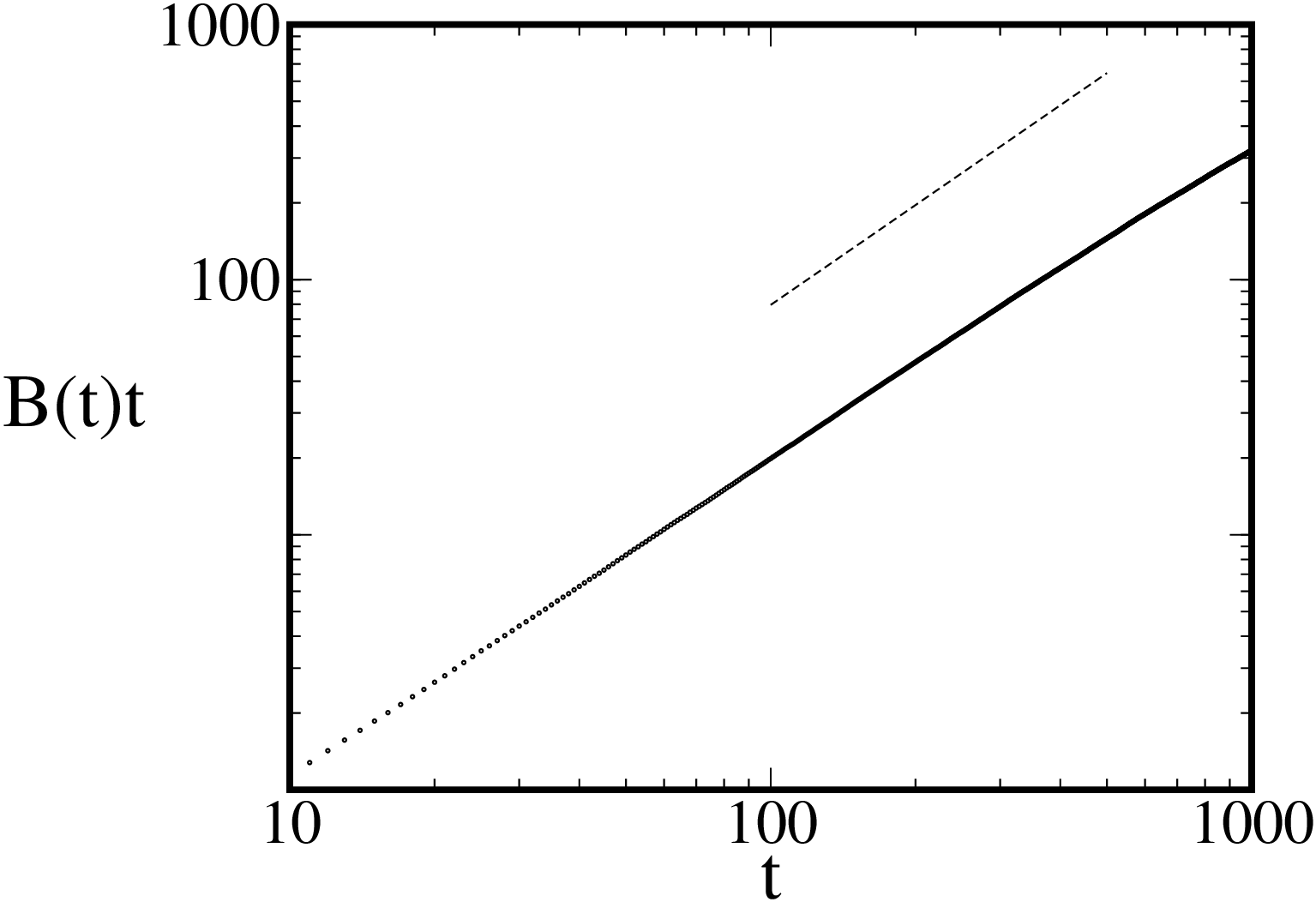}
\end{center}
\caption{$B(t)t$ as a function of $t$ in the case $\bar\rho=0.5$ 
and $L=128$.  The  guided  line (the dashed line) represents   
$B(t)t\propto t^{4/3}$. 
}
\label{fig:g}
\end{figure}

The non-exponential decaying behavior of $C(t)$ may be related to a 
long time tail of  the  time  correlation  function  of  current 
fluctuations.  Indeed, using  the  continuity  equation,  $C(t)$  
is expressed by
\begin{equation}
C(t)=\left( \frac{2 \pi}{L} \right)^2 
\frac{1}{2} \bra
\left|\sum_{k=1}^{L\tau} \sum_{x=1}^L j_x(\eta(k-1)\rightarrow\eta(k)) 
\e^{i \frac{2\pi}{L} x} \right|^2  
\kets{E}
\end{equation}
It is  expected  that  the  behavior  of $C(t)$ in a shorter  time 
regime than the diffusion time $D^{-1} (2\pi/L)^{-2}$  has  common 
features with that of 
\begin{equation}
B(t)=\frac{1}{2} \bra 
\left|\sum_{k=1}^{L t} \sum_{x=1}^L \left[j_x(\eta(k-1)\rightarrow
\eta(k)) -\bar J_E\right] \right|^2  \kets{E},
\end{equation}
which is displayed in   Fig. \ref{fig:g} in the case $\bar\rho=0.5$ 
and $L=128$. The graph is consistent  with the relation $B(t)t \propto 
t^{4/3}$ which was derived from the theoretical analysis of 
fluctuating hydrodynamics \cite{BKS}. 

We remark that these observations on $R(t)$ and $C(t)$ cannot   be    
explained  from the assumption that  the time evolution of density 
field is described by a linear Langevin equation of 
$\hat \eta_1 (t)$, because under this assumption both the relations  
(\ref{ein1})  and  (\ref{frr2})  are derived though (\ref{frr2}) was 
found not to valid. This implies that the dynamical aspect of the 
system is too complicated. It may be surprising that  the 
extended Einstein relation holds with the appropriate choice of $D$. 
 
%%% new relation %%%

\begin{figure}
\begin{center}
\includegraphics[width=8cm]{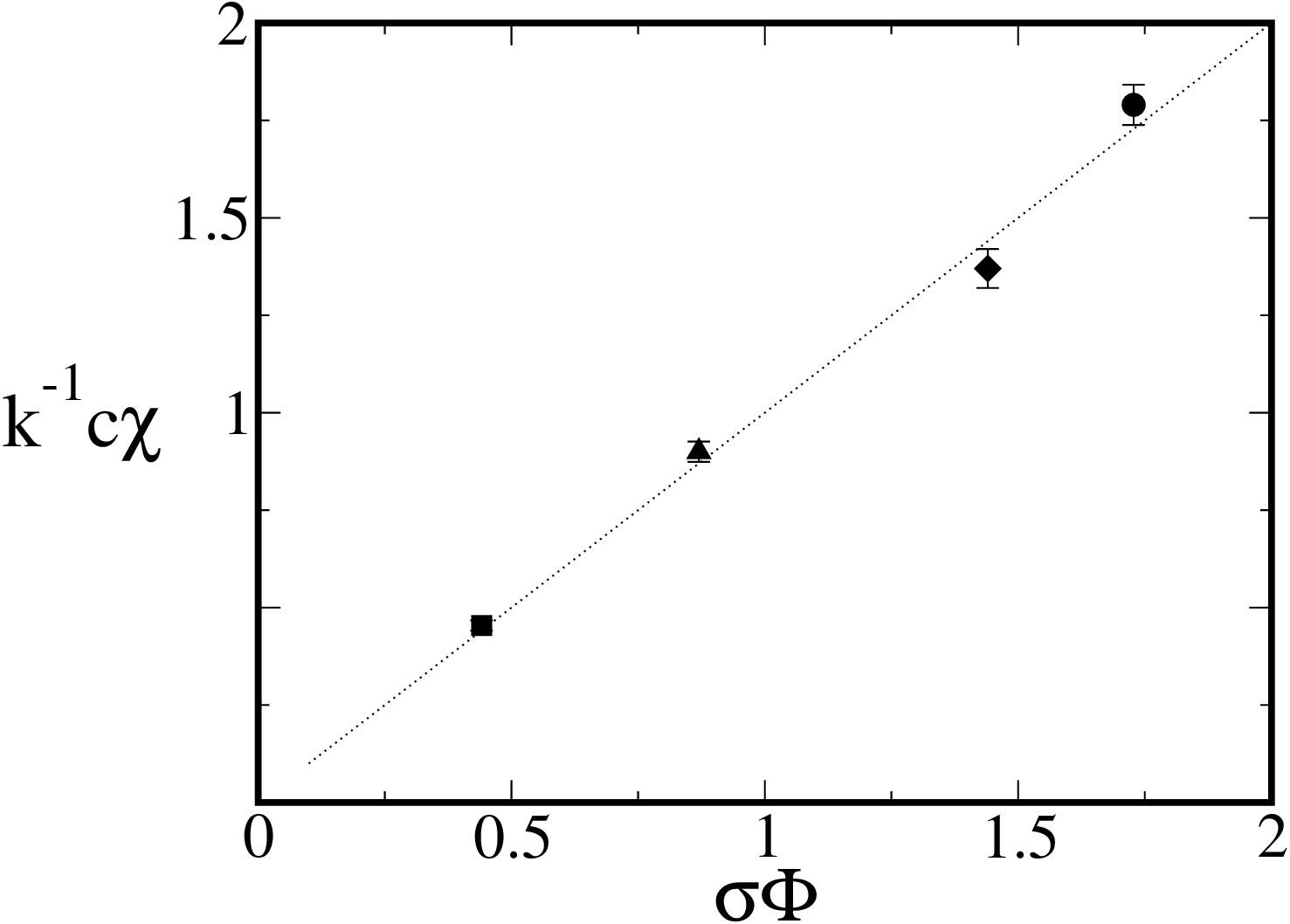}
\end{center}
\caption{$k^{-1}c\chi$  as a function of $\sigma\Phi$, where 
$k=2\pi/L$, for the case $\bar\rho=19/64$. Note that for the 
case $\bar\rho=0.5$, $c=\Phi=0$. The  cases  $L=64, 128, 192$  and  
$256$ correspond to the square, triangle, diamond  and circle, 
respectively.  The straight line represents $(L/2\pi)c\chi=\sigma\Phi$.  
The error bars represents the standard deviations of the statistical 
samples. In addition to the statistical error bars, there exists a 
fitting error which is less than 5 percent of the quantities. 
}
\label{fig:ein2}
\end{figure}

Finally, from (\ref{ein33}),  we find a new relation 
\begin{equation}
\frac{L}{2\pi}c(L)\chi(L)=\sigma(L)\Phi(L).  
\label{ein3}
\end{equation}
We investigated the  validity of (\ref{ein3}) in numerical experiments.  
We numerically measured $c$  from  the oscillatory behavior of the time 
correlation function (\ref{corre}). In  one example of this  analysis, 
we considered all $t_i$ satisfying  $C(t_i)-C(\infty)=0$, 
where $t_{i-1} < t_i$. Then, $c$ can be estimated from the relation
\begin{equation}
c \frac{2\pi}{L} (t_{i+1}-t_i)=\pi .
\end{equation}
We have confirmed that $c$ converges to $c_*=0.17$ in the case 
$\bar\rho=19/64$ in the limit $L\to\infty$. In Fig. \ref{fig:ein2}, 
we plot $(L/2\pi)c(L)\chi(L)$ as a function of  $\sigma(L)\Phi(L)$ 
for the cases $\bar\rho=19/64$, with $L=64,128$ and $256$. (Note 
that $c=\Phi=0$ for the case $\bar \rho=0.5$.)  This graph clearly 
shows that the extended  Einstein relation (\ref{ein3}) holds within 
the precision of the numerical computations.

%

%%%
%%%
\section{Thermodynamic consideration}
%%%
%%%  

%%%  thermodynamics  %%%

Here, we devote space to thermodynamic consideration related to 
our results. We first study the equilibrium case $E=0$ by 
considering a situation  that  a perturbation  potential is 
applied to the equilibrium system.  
When we choose a slowly varying potential, we can regard $V_x$ and 
$\bra\eta_x\kets{E=0,V}$ as smooth quantities $V(x)$ and $\rho(x)$. 
By use of the chemical potential $\mu(\rho,T)$ of the equilibrium 
system (without the potential $V(x)$),  a density profile under the 
influence of $V(x)$ is determined by a relation that 
\begin{equation}
\mu(\rho(x),T)+V(x)={\rm const. \hspace{2mm} in \hspace{2mm}}  x,  
\label{defmu}
\end{equation}
 which is equivalent to   
\begin{equation}
\left(\pder{\mu}{\rho}\right)_T \der{\rho}{x}+\der{V}{x}=0. 
\label{thermo:1}
\end{equation}
Note that the fluctuation relation 
\begin{equation}
\left(\pder{\mu}{\rho}\right)^{-1}_T=\frac{\chi}{T}
\label{fr}
\end{equation}
holds in the equilibrium state.  As another relation in this situation,  
we have the total zero current condition 
\begin{equation} 
-D\frac{d\rho}{dx}+\bar{J}\left(\rho,-\frac{dV}{dx}\right)=0. 
\label{thermo:2}  
\end{equation}
By expanding the second  term of (\ref{thermo:2}) in $dV/dx$, we obtain  
\begin{equation}
D\der{\rho}{x}+\sigma \der{V}{x}=0. 
\label{thermo:3}
\end{equation} 
A combination of  (\ref{thermo:1}), (\ref{fr}) and (\ref{thermo:3}) 
leads to the Einstein relation (\ref{eqein}).  
In the equilibrium case, we can show just the opposite 
that a combination of the Einstein relation (\ref{eqein}) and  the 
relation (\ref{thermo:3}) derives  the fluctuation relation 
(\ref{fr}) through the relation (\ref{defmu}).    

In connection with the above mentioned thermodynamic consideration,  
we introduce our previous study  on a two-dimensional DLG \cite{HSI}.   
First, we remark that an appropriate definition of the chemical 
potential for a NESS  is not known, unless we use the local 
equilibrium assumption. Because in NESS far from equilibrium,  
the local equilibrium assumption  is not valid,  we attempted to 
define a chemical potential without this assumption.   Focusing on 
a density response in a direction perpendicular to the driving force, 
we defined a nonequilibrium chemical potential by 
a similar relation with (\ref{defmu}) and constructed an extended 
thermodynamic 
function from  measurable quantities. As a non-trivial prediction, 
we showed numerically that the rate function of density fluctuation 
was determined by this extended thermodynamic  function.  A theoretical  
framework behind these observations has been uncovered 
\cite{SasaTasaki}.       We remark that the Einstein 
relation (\ref{eqein}) was also observed  in a direction 
perpendicular to the driving force \cite{Hayashi} and 
that the argument in the previous paragraph can be applied to this 
case.  

In the present paper, we have reported that the extended Einstein 
relation (\ref{ein1}) and the new relation (\ref{ein3}) hold in NESS, 
and  a combination 
of both the relations and (\ref{ein22}) that corresponds to the Fourier 
transform of (\ref{thermo:3}) leads to  the definition of the 
effective temperature (\ref{ctheta}). Because in the case $E=0$, 
the relation (\ref{eqeq}) is derived from (\ref{thermo:1}) and (\ref{fr}), 
one might expect that  there exists a nonequilibrium chemical 
potential even for the present NESS. For example,  
as a trial, one might propose a nonequilibrium chemical potential   
$\hat\mu_1(\hat\rho_1,\hat\theta_1)$ as   
\begin{equation}
\hat\mu_1(\hat\rho_1,\hat\theta_1)+\hat  V_1=0.    
\label{muk}  
\end{equation}
However, (\ref{muk}) and (\ref{ctheta}) lead to 
\begin{equation}
\left(\frac{\hat\mu_1}{\hat\rho_1}\right)^{-1}=
\frac{\chi}{\hat\theta_1}.    
\label{fr2}
\end{equation}
Because the relation (\ref{fr2}) is hard to be interpreted as 
an  extended form of the fluctuation relation (\ref{fr}),  the 
definition  (\ref{muk}) seems not to be useful.  Taking into 
consideration that the relation (\ref{muk}) is just one  
trial,  there might be a plausible definition of a chemical potential 
for NESS, but we have not found it yet. The difficulty that does not  
arise in the equilibrium  case lies in  the wavenumber dependence 
of $D$, as is expected from the system 
size dependence of $D$ shown in Fig. \ref{fig:dif}.  
This wavenumber dependence  is not observed in the equilibrium 
case and prevents us from pushing ahead further thermodynamic 
consideration.

\begin{figure}
\begin{center}
\includegraphics[width=8cm]{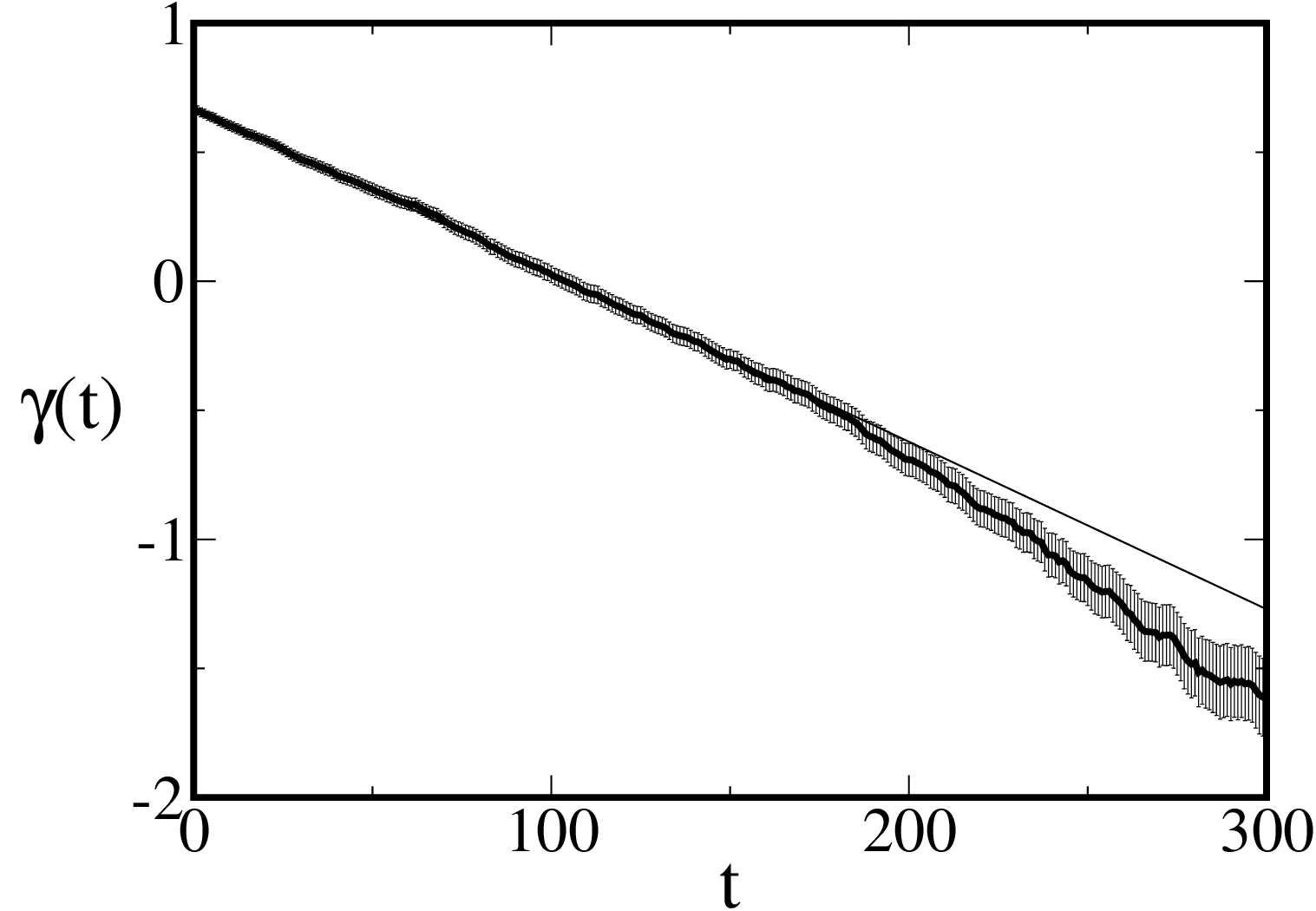}
\end{center}
\caption{In the case $H_0(\eta)=0$ (ASEP), 
$\gamma(t)$  is plotted  with error bars in the case 
$\bar\rho=0.5$, $L=64$ and $|\hat V_1|/\sqrt{L}=0.1$ 
(see (\ref{d:diff})). The slope of the  fitted  line  
(thin line)  in  the  early  stage ($t <  150$ mcs) 
corresponds to  $D(2\pi/L)^2$. 
}  
\label{fig:asep1}
\end{figure}

%%%
%%%
\section{Discussion}
%%%
%%%  

Equations (\ref{ein1}) and (\ref{ein3}) constitute the main results 
of this paper. In the following,  we give five important remarks that 
are peripherally related to  these main results.

%%% effective temperature %%%

First, we give an argument from which to understand 
a physical meaning of $\hat\theta_1$ as the effective temperature. 
Effective temperatures have been studied mainly in glassy systems 
\cite{CKP,BB,DVL,jamm,Harada}. The simplest characterization of 
temperature may be provided by the equi-partition. The law of 
equi-partition is regarded as one example of an energy distribution 
law that can  be expressed by the canonical distribution. Out of 
equilibrium, the equi-partition law and the canonical distribution 
do not  hold  in  general.  Indeed,  the  steady  distribution 
$\ps{E}(\eta)$ differs greatly  from the canonical   one. However, 
as  demonstrated  in  Refs. \cite{BB,HSIIIa},  by  constructing 
measurement methods to extract  behavior at some large time and/or 
length scales, the equi-partition law and the canonical distribution 
are recovered at these scales.  

Hypothesizing a similar situation for  the NESS we study, we  
assume a simple form for the modification of the steady state 
measure under the condition that the potential varying slowly in 
space is added to the system. The form we consider is
\begin{equation}
\ps{E,V}(\eta)= \frac{1}{Z_V} \ps{E}(\eta) 
\e^{-(\hat \eta_{-1} \hat \theta_1^{-1} \hat V_1+\rm c.c.)}
\label{wh}
\end{equation}
for $|\hat V_1| \ll 1 $, where  $Z_V$ is a normalization constant.     
{}From (\ref{wh}), we can derive (\ref{ctheta}),  which is the 
definition  of  $\hat\theta_1$.  We believe  that the simple form 
(\ref{wh}), which  describes the response  of  the  steady  state 
measure to  this  slowly varying potential, is a good starting point 
to investigate NESS.

%%%  asep %%%

Second, we remark that our main result also holds for the case
$H_0(\eta)=0$, which is an  example of asymmetric simple exclusion 
process (ASEP). Specifically, we consider the case $\rho=0.5$ and 
$L=64$. $\chi$ and $\sigma$  are  calculated  theoretically  as 
$\chi(L)=0.25$ and $\sigma(L)=0.043$. The diffusion constant $D(L)$ 
is measured numerically as $D(L)=0.68$ (see. Fig. \ref{fig:asep1}). 
$\Theta$ is also measured as $\Theta(L)=3.9$ from (\ref{ctheta}).  
Thus  the  relation $D(L)\chi(L)=\sigma(L)\Theta(L)$ seems valid. 
Study on the large scale behavior of the system might give new 
insights on ASEP.  

% 

%%%  fh  %%%  

Third, we remark a connection with fluctuating hydrodynamics. 
Note that the system in our numerical experiments may be too small 
for the dynamical behavior to be described within the framework of 
fluctuating hydrodynamics, while large enough so that we can check 
universal relations around equilibrium states. The fact that 
$D(L)\propto \sqrt{L}$ (See Fig \ref{fig:dif}) in such small systems 
seems to imply that bare parameters appearing in fluctuating 
hydrodynamics depend on a length scale $\ell$ at which hydrodynamic 
description is defined. (Note that the scaling relation $D(L)\propto 
\sqrt{L}$ can be derived within the framework of fluctuating 
hydrodynamics, irrespective of values of the bare parameters 
\cite{BKS}.)  Thus, it is plausible to conjecture that we need to 
specify  a scale $\ell$ explicitly so as to have a hydrodynamic 
description.  We do not understand how the formal derivation in 
Ref. \cite{eyink} corresponds to this observation.  

%%% tracer diffusion %%%

Forth,  we give a comment on  diffusion of a marked particle.  
The diffusion is characterized by a tracer diffusion constant,
which is not equal to the bulk diffusion constant, $D$, defined 
by (\ref{d:diff}). One may has a question whether the tracer 
diffusion constant can be involved in a universal relation for NESS. 
We do not find such a relation yet, but it may be an interesting 
future problem to seek it. We note that the tracer diffusion constant 
does not have the anomalous $L$ dependence in contrast to the case 
of $D$.

%%% experiments %%%

The final remark  we wish to present regards the  possibility 
of an experiment that can  check  the extended Einstein relations
we have proposed.  We conjecture that it may be best to study a 
system in which  colloid particles, confined to  one-dimension, are 
driven through a periodic array of optical traps \cite{KTG}. By 
defining the effective temperature in the same  way  as  did  in 
(\ref{ctheta}), the validity of the extended Einstein relations 
will be checked.

The authors acknowledge H. Tasaki, T. Harada and M. Sano for stimulating 
discussions on NESS. This work was supported by a grant from the Ministry 
of Education, Science, Sports and Culture of Japan (No. 14654064). 

%%%%%%%%%%%%%%%%%%%%
%%%% references %%%%
%%%%%%%%%%%%%%%%%%%%


\begin{thebibliography}{99}

%%%% introduction %%%%%%%%%%%%%%%%%

\bibitem{ein}
A. Eintein, Annalen der Physik {\bf 17} 549, (1905); 
\begin{em} Investigations on the theory  of Brownian movement\end{em},
(Dover, 1956).

\bibitem{ein2}
R. Kubo, Rep. Prog. Theor. Phys. {\bf 29} 255, (1966).

\bibitem{HSIIIa}
K. Hayashi and S.  Sasa, Phys. Rev. E {\bf 69}, 066119, (2004).     

\bibitem{KLS}
S. Katz, J. L. Lebowitz and H. Spohn, J. Stat. Phys. {\bf 34} 497, (1984).

\bibitem{BKS}
H. van Beijeren, R. Kutner, and H. Spohn, Phy. Rev. Lett. {\bf 54} 2026, 
(1985).

\bibitem{SP}
H. Spohn, 
\begin{em} Large Scale Dynamics of Interacting Particles\end{em} ,
(Springer-Verlag, 1991).  

\bibitem{HSIVb} K. Hayashi and S. Sasa, in preparation 

\bibitem{ER}  There is a puzzle that the extended Einstein relation 
seems to connect this $D$ with $\sigma$, because  $D$ is evaluated 
form the short time behavior of the system,  while $\sigma$ is 
measured as the long time average in the steady state. 



% --  effective temperature  --

\bibitem{CKP} L.F. Cugliandolo, J. Kurchan and L. Peliti, Phys. Rev. E 
{\bf 55}, 3898, (1997). 

\bibitem{BB} L. Berthier and J-L Barrat, Phys. Rev. Lett., {\bf 89}, 
095702, (2002). 

\bibitem{DVL} A. B. Kolton, R. Exartier, L. F. Cugliandolo, 
D. Dominguez and N. Gronbech-Jensen,  Phys. Rev. Lett., {\bf 89}, 
227001, (2002). 

\bibitem{jamm} I. K.  Ono et al, Phys. Rev. Lett., {\bf 89}, 095703, (2002). 

%\bibitem{Cri} A. Crisanti and F. Ritort, cond-mat/0212490; check J. Phys. A  
%specieal issue. (2003). 

\bibitem{Harada} T. Harada, cond-mat/0310547, Europhysics Letters, 
in press.  

\bibitem{HSI} K. Hayashi and S. Sasa, Phy. Rev. E 68, 035104(R) (2003).  
\bibitem{SasaTasaki} S. Sasa and H. Tasaki, cond-mat/0411052. 
\bibitem{Hayashi} K. Hayashi, in preparation.


%%%

\bibitem{eyink}
G. L. Eyink, J. L. Lebowitz, and H. Spohn, J. Stat. Phys. {\bf 83} 
385, (1996) and references therein..  

\bibitem{KTG}
P. T. Korda, M. B. Taylor, and D. G. Grier, Phys. Rev. Lett. {\bf 89}  
128301, (2002).  


\end{thebibliography}
\end{document}